\documentclass[12pt]{article}
   
   \usepackage{amssymb}
   
\newcommand{\BEQ}{\begin{equation}}     
\newcommand{\BEA}{\begin{eqnarray}}
\newcommand{\EEQ}{\end{equation}}       
\newcommand{\EEA}{\end{eqnarray}}
\newcommand{\eps}{\varepsilon}          % epsilon

\begin{document}
\begin{titlepage}
\noindent 

 \begin{center}
{\Large \bf Infinite dimensional  covariance and non relativistic limits 
  in time dependent theories }
\end{center}

 \vskip 2.cm
 \centerline{  {\bf Antoine D.  Coste} \footnote{53 rue des meuniers, 
 F 92150 Suresnes,  France}  }
 \vskip 0.5 cm
 
\centerline {Laboratoire de Physique 
 Th\'eorique,\footnote{ CNRS UMR 8627}
 Campus de l' Universit\'e de Paris-Sud, France} 

%  \centerline{  F -- 91405 Orsay Cedex, France}
 
%B\^atiment 210,

 \vskip 2.cm 
 \begin{abstract}
 We give here some account of investigations for the possible role of 
  infinite dimensional Lie algebras, whose  simplest example is the classical 
   Virasoro algebra, in time dependent systems or anisotropic statistical 
    models.  We expect a central extension to arise due to quantum corrections, 
    but we first  define the classical objects and 
    so called "primary fields"  transformation laws  in order to be able to identify 
    it precisely.  The issue of non relativistic limit (negligible Compton wave 
    length , or very big speed of light ) 
    of Minkovskian field theories is also of importance here.  
 \end{abstract}

\end{titlepage}

 \section{ Introduction }
 \par Conformal theories, that is physical theories based on general relativity 
   intuition ( covariance laws, energy momentum tensor, differential symmetries), 
   and using mathematical constructions such as Virasoro algebra and 
   representation theory of infinite dimensional algebra, have a great predictive power 
   in modern description of critical two-dimensional phenomena. 
   Such domains not only include 
   statistical physics 2d models such as  Ising and Potts models, but also 
   quantum spin chains, or effective formulations such as rotationally invariant 
   quantum theory around Kondo effect impurities, magnetic monopoles or 
   black holes. \\ 
   Very soon  attempts have been made to extend these ideas and concepts 
   such as central charge to a wider 
   physical context, such as higher dimensional field theories at some 
   renormalisation group fixed point. \\   
    Good news may come from another front, where in tackling anisotropic 
     statistical models, or time dependent problems, M. Henkel 
     recognized  some classical Virasoro symmetry and infinite dimensional 
     generalisation of Galileo transformation. \\  
     It is also an important problem to deal with a complete description 
     of renormalisation group flow between various critical theories, 
     using eventually operators  similar to heat kernels. \\  
 We present here some first attempts   in this new program, hoping to give 
 more thorough computations, results and physical 
 applications in the future. These include  issues such as classical 
 limit of relativistic theories, a topic discussed in a rather difficult  
 paper by Barut. 
 %%%%%%%%%%%%
\section{ General set up for differential operators } 
{\bf  Proposition}

\par On any d-manifold, if $x$ is a chart, 
$ B = \sum_{\mu =1 }^{d} \, B^{\mu }(x) \partial_{\mu } $\  is a non vanishing, 
 smooth vector field, $C=C(x) $  is an integrable  function on any smooth path, 
  $\rho $ is a small real number, then \   
\BEQ 
\phi (x, \rho  ) =  exp ( \rho ( B + C )  ) \ \psi (x) \   
                 =\ exp  \Big( {\cal T}( x, \rho )  \Big) \  \psi (x'(x,\rho ))
\EEQ 
where the local diffeomorphism $x' $ doesn't depend on $C$, 
and ${\cal T}$ is explicitly given below.\\  
{\bf Proof } is geometric:\\    
   We define $x'(x,\rho ) $ to be the flow of the vector field $B$, i.e. 
( in an appropriate neighbourhood of the point of coordinates $x$ ) 
the solution of 
\BEQ  \frac{ \ \ \partial x'^{\mu } }{\partial {\rho } }  = B^{\mu } (x'(x,\rho ))
\EEQ  
   satisfying the initial condition $ x'(x,\rho =0) =  x $. 
   On the other hand, definition 
   of the $ {\cal T} $ "phase" is: 
  \BEQ  {\cal T}(x, \rho  )= \int^{\rho  }_{0} \  C(x'(x,s)) \   ds 
   \EEQ 
$   \phi $ is identified as the solution of  
\BEQ  \frac{ \partial  \phi }{ \partial \rho } =  ( B + C )  \phi  
\EEQ  
satisfying the initial condition $ \phi (x, 0) =  \psi (x) $. Differentiation 
with respect to $\rho $ gives 
$$  lhs =  ( B+C) (x') \phi  $$ 
whereas application of $(B+C)(x)$ to the r.h.s. gives an apparently different 
expression. 
In fact  proving the above proposition amounts to 
 establishing   the \\  
%%%%%%%%
   {\bf Key lemma: } for any $\nu  $, and small enough  $\rho $  : 
\BEQ  B^{\nu }(x'(x, \rho ))  = \sum_{\mu } B^{\mu }(x) \    
\frac{ \partial x'^{\nu } }{\partial x^{\mu } } (x, \rho  )
\EEQ  

(This means the solution $x' $ is such that B at the image is the  transform of 
B at the origin , in the sense of change of coordinate systems formulae....) \\  
  
   {\bf Proof } of the lemma:  \\ 
   Let H be an hypersurface containing $x $ and such that $B(x) $ is  not tangent to H. 
   We restrict ourself to a domain of H containing $x$ where $B(h)$ is not tangent to H.
   Then the map  : 
\BEA   
   H \times \mathbb{R} &\rightarrow & \mathbb{R}^d   \nonumber \\    
   (h, \delta )   &\rightarrow &  y(h, \delta ) \ \ \hbox{  solution of } \ \ 
   \frac{ \partial y^{\mu } }{ \partial \delta  } = 
         B^{\mu }   ( y(h,\delta ) )   \nonumber \\  
  & &  \hbox{ with initial  condition } \ \ y(h,\delta =0) = h  \nonumber 
\EEA    
      
  is one to one in a neighbourhood of $x $ , and its inverse 
   $x \rightarrow (h_x ,\delta_x )$
  allows us to define instead of $x^{\mu }$ a new coordinate system $h^\alpha $ :
  \BEQ  h^0 =\delta \ ,\hbox{and } \  (h^{\alpha }), \ \alpha = 1,\dots ,d-1 
  \ \hbox{are local coordinates on H } 
   \EEQ 
   This system is such that  $ x'(x,\rho ) = y( h_x , \rho + \delta_x ) $ 
   for $x$ and $x'$ close  enough to $H$. \\   
%%%%%%%%
   A simple reasonning expressing Jacobian matrices between these two 
    coordinate systems gives the lemma: $ J_1 = \frac{Dy^{\nu }}{Dh^{\beta } } $
$$    \hbox{has line }\nu \hbox{ made of } \  \Big( B^{\nu } (y) , \ 
    \frac{\partial y^{\nu } }{\partial h^{\beta } } \  , \   
     \beta = 1,\dots ,d-1 \Big)   $$
Its inverse       $ J_2 = \frac{Dh^{\alpha  }}{Dx^{\nu  } } $ has line 
 $\alpha =0  $ equal to $ \frac{\partial \delta }{\partial x^{\nu }}  $ \\    
 and other elements equal to  
$ \frac{\partial h^{\alpha  } }{\partial x^{\nu } } $.\\  
\BEA  
 \hbox{Therefore  } \   1 &=& (J_2 J_1 )^0_{\ \  0} = \sum_{\nu =1 }^{d} \    
 \frac{\partial  \delta }{\partial x^{\nu } } (y) \,       B^{\nu } (y)\nonumber  \\  
 0   &=& (J_2 J_1 )^{\alpha }_{\ \ 0} = \sum_{\nu =1 }^{d} \  
 \frac{\partial h^{\alpha  } }{\partial x^{\nu } } (y) \,  B^{\nu } (y)\nonumber 
\EEA  
which we can use at $ y=y(h_x , \delta_x ) =\,  x $,  so that: 
\BEA 
& &  \sum_{\nu =1 }^{d} \   B^{\nu } (x)\, 
 \frac{\partial x'^{\mu  } }{\partial x^{\nu } \ |_{\rho } } 
  = \sum_{\nu =1 }^{d} \   B^{\nu } (x)\, \frac{\partial }{\partial x^{\nu } }
\Big(  y^{\mu } (h_x \,  , \rho + \delta_x  ) \Big)          \nonumber \\   
& =&  \sum_{\nu =1 }^{d} \   B^{\nu } (x)\,  \Big(   B^{\mu } (x'(x,\rho )) 
    \frac{\partial  \delta_x }{\partial x^{\nu } } +  
    \sum_{\alpha =1 }^{d-1} \  
    \frac{\partial  y^{\mu }(h,\rho +\delta ) }{\partial h^{\alpha  } }\,  
    \frac{\partial h^{\alpha  }(x) }{\partial x^{\nu } }\   \Big) \nonumber \\ 
& =&  B^{\mu } (x')\, (J_2 J_1 )^0_0  +  \sum_{\alpha =1 }^{d-1} \  
  \frac{\partial  y^{\mu } }{\partial h^{\alpha  } }\,  
  (J_2 J_1 )^{\alpha }_0    \nonumber \\ 
& =&  B^{\mu } (x')         \nonumber  
\EEA 
%      \frac{\partial  }{\partial } 
 %%%%%%%%%%%%%%%%%%
 \vskip .5cm   
%%%%%%%%%%%%
Another (direct) proof is possible by systematic use of Dirac distributions  and Green's 
functions. For any differential operator   $O(x) $ : 
\BEA \Big( e^{\rho \, O(x) }\ \psi  \Big)(x) &=&  
 e^{ \rho \, O(x) } \int \, dy \  \delta (x-y) \psi (y) \nonumber  \\ 
 &=& \int \, dy\   G_{\rho } (x,y) \psi (y) \\  
 \hbox{where } G_{\rho } (x,y) &=&  e^{\rho  \, O(x) } \  \delta (x-y) 
\EEA  
 
 Therefore the above proposition amounts to: 
\BEA 
 G_{\rho } (x,y) &=& \Big(  exp \int_0^{\rho } \ C( x'(x,s) )\, ds \Big) \ 
                          \delta ( y-x'(x, \rho ) )    \\  
                 &=&  \sum_n  \frac{\rho^n  }{n! } \ 
   \Big( C(x)+ B^{\mu }  \,
          \frac{ \partial  }{ \partial x^{\mu } } \Big)^n \   
            \delta ( x-y )  
\EEA   
$    \delta ( y-x'(x, \rho ) )  $ can be computed by use of:
\BEQ  A^{\mu }(x):= x'^{\mu } (x, \rho )-x^{\mu } \ = \sum_{n=1 }^{\infty} \  
 \frac{\rho^n  }{n! } \  \Big( (B\cdot  \partial_x )^{n-1} B^{\mu } \Big) (x) 
\EEQ 
\BEA 
&&\hbox{Thus } \ \delta ( y-x'(x, \rho ) )   \\  
 &&= \ \delta ( y- A^{\mu }(x ) -x) = \ 
 exp\Big( - \, A^{\mu }(x ) \frac{ \partial  }{ \partial y^{\mu } } \Big)  
       \     \delta (y-x)  \nonumber \\  
 &&=\Big( 1+ \rho B^{\mu }    \partial_{\mu }  
        + \frac{ \rho^2 }{2} \Big( (B\cdot \partial B^{\mu } )
          { \partial_{\mu } } + 
          B^{\mu } B^{\nu } \partial_{\mu} \partial_{\nu}\Big)  +\dots   
     \Big) \delta (x-y)    
\EEA 
$$\hbox{where  }\  \partial_{\mu} = {\partial  \over  \partial x^{\mu } } $$  

%%%%%%%%%%%%%%%%%%%%%
 \section{ An interpretation of  formulae  for non relativistic limits} 
%%%%%%%%%%%
\par We present here a geometric interpretation of a paper by Barut\cite{Baru73} 
 relating   non relativistic limits and group contraction. 
 A peculiar case of the  
 above proposition  reads:  
\BEA 
\phi &= & exp \Big( \rho  f(t) \Big( \frac{ 2\pi \, mc^2 }{ h} +
\partial_t \Big) \Big)    \ \psi (t)  \nonumber \\  
&= &  exp\ \Big( \frac{  2\pi \, mc^2 }{ h}  \Big( t'(t,\rho )-t \Big)   \Big) 
 \psi ( t'(t,\rho ) )
\EEA 
A direct proof of this equation, can be easily set up by considering the function 
\BEA
 t\ &:=& \  T(\delta )  \ \ 
\hbox{reciprocal  of } \ \  \delta \ := \int^t \frac{ du}{f(u) }  \\   
\hbox{ then } \ \ t' &:=& T(\delta + \rho ) = \ t'(t,\rho )          \\   
\hbox{and } \  \rho  &= & \int^{t'}_{t} \;  \frac{ du}{f(u) } \ \  ,\ \   
  d\rho = \frac{ dt'}{ f(t')}\ - \frac{ dt}{ f(t)} 
\EEA 
%%%%%%%%%%%%%%%%%%%%%%%%%%%%%%%%%%%%%%%%%%%%%%%%%%%%%%
\par  
We would like to point out a rigorous geometric way of reproducing 
  some of Barut's results: set 
\BEA   x_0 := c\tau \ ,\  \phi_t  ( x_0 , x)\ &:=& 
  \ exp \Big ( \tau \Big ( \frac{  2\pi \, mc^2 }{ h} +  \partial_t \Big)   \Big)
   \ \psi (t,x)   \\  
        &=&  exp \Big ( \frac{  2\pi \, mc^2 \tau }{ h}\Big)  \  \psi (t+\tau ,x) 
  \EEA
  This functional relation  
   is such that flat space Klein-Gordon operator applied to $\phi ( x_0 , x) $    
 is proportional in the non relativistic limit to diffusion one applied to 
 $\psi (t,x) $. More precisely define: \\   
\BEQ \Psi_{x_0 } (t,x) := exp \Big ( \frac{  2\pi \, mc x_0 }{ h} \Big) 
   \psi \Big( t+ \frac{x_0 }{c}\  ,x \Big)  =  \phi_t  ( x_0 , x) 
 \EEQ 
If we consider in $ \Psi_{x_0 } (t,x)  $   $x_0  $ as a parameter, and $t$ as a parameter 
 in $ \phi_t  ( x_0 , x)  $ , we have:  
\BEQ 
  \frac{\partial }{ \partial x_0 } \phi_t  = 
  \Big( \frac{ 2\pi \, mc}{h} + \frac{1}{c} \frac{\partial }{\partial t} \Big)
 \Psi_{x_0 } 
 \EEQ  
This identity expresses the so called "group contraction trick"  which insures:  
 \BEA 
  \Big(\partial_{0}^{2} -  \partial_{x}^{2} - 
   \frac{4 \pi^2  m^2 c^2 }{h^2 } \Big)& &  \phi_t  ( x_0 , x)  =  \nonumber \\  
 & &  \Big( \frac{4\pi m}{h}  \partial_{t} - \partial_{x}^{2} +
    \frac{1}{c^2 }   \partial_{t}^{2} \Big) \  \Psi_{x_0 } (t,x)  
  \EEA 
 The last term of the r.h.s.  being negligible in the following 
 "non relativistic limit".
 
%%%%%%%%%%%%%%%%%%%%%
\section{  Space - Time  covariance} 
%%%%%%%%%%%%%%%%%%%%%%%%%%
 \par The following differential operators, considered by M. Henkel in his book, 
\BEA  -X_{\eps  } \ &:= &\  \eps (t) \partial_t 
      + \frac{N}{2 } {\dot \eps} (t) \ 
       (\ r\partial_r  +\chi )
     + \frac{mr^{2/N}}{4 } {\ddot \eps } (t)      \\    
      &:=& - \sum_{n\in \mathbb{Z} } \epsilon_n X_n   \\ 
 \hbox{with  } \ \ \   \eps(t)  &:= & \sum_{n\in \mathbb{Z} } 
 \eps_n \, t^{n+1}   
\EEA   
satisfy classical Virasoro algebra: 
\BEQ   [ X_{\eps } , X_{\eta } ] =  
         X_{ {\bf {\dot \eps }\ \eta - \eps  {\dot \eta } } } 
\EEQ  
It is therefore a challenge to try using Virasoro covariance in the context of 
 time dependent physics such as heat diffusion . A first step is derivation 
 of some "primary field" transformation law:   \\  
 
 {\bf Proposition } 
 \BEA  
& & \phi (t,r, \eps ) := exp (-X_{\eps } ) \psi (t,r)  \\  
 &  &= \Big( \frac{ \eps (t')}{\eps (t) }\Big)^{ N\chi /2 }
    \ exp \Big( \frac{m }{4}\, \Big( 
     \frac{r'^{\ 2/N} \, {\bf \dot \eps }(t')}{\eps (t')} - 
     \frac{r^{   2/N} \, {\bf \dot \eps }(t )}{\eps (t )}  \Big) \Big) \ 
    \psi (t'\, ,r')  \\  
& & r'= r'(t,r, \eps ) 
   = r\   \Big( \frac{\eps (t')}{ \eps (t) } \Big)^{N/2}  \\ 
 t'= t'(t,\eps ) & & \hbox{ such that }\   \  \ 
        1= \int_t^{t'} \frac{d\tau }{ \eps (\tau )}  
\EEA   
{\bf Proof}  of this law is obtained following the lines of the general method given in 
 first section, note that t' here does not depend on $r,\  m,\  \chi , N $   
 and that  
 \BEQ 
 {\bf {\ddot \eps  }}(t'(\rho ))\  {\eps (t'(\rho ))} = 
 \frac{ d}{d\rho }\,  {\bf {\dot \eps }}(t'(\rho )) \  \ . 
\EEQ  

Note that if we introduce some time dependent scale $\sigma $ 
\BEQ    
 \eps (t )\ := e^{ \sigma (t) }  
\EEQ 
\BEA  \phi(t,r)   &dt^{\ n\chi /2 } & \  exp \Big( \frac{ mr^{2/n  }}{4} \,  
    {\bf {\dot \sigma  }} (t) \Big)  \nonumber \\  
  =\  \psi(t',r') &dt'^{\ n\chi /2} & \  exp \Big( \frac{mr'^{\ 2/n}}{4} \,  
    {\bf { \dot \sigma }} (t')               \Big) 
\EEA  
From now on we take $N = 2/\theta  =1 $ in the language of 
statistical mechanics, this 
corresponds to the heat kernel situation.   
 
\par  A second step is to derive correlation in geometries related by such transformations: 
as an academic example suppose a two parameter diffeomorpism between 
flat space $(t,r) \in \mathbb{R}^d $ and half space: 
\BEQ  t' = T' \ exp ( t/T ) 
\ \ \ \ \     \frac{r'}{r} = \sqrt{ \frac{T'}{T}  } \ exp( t/2T \, ) 
\EEQ   
If $ \phi (t,r) $ has propagator of a massless scalar of dimension  $(d-2)/2 $ we will 
obtain the prediction for:
\BEA  0<t' <\infty \ & & \   < \psi (t',r' ) \psi(0,0) > \ =   \nonumber   \\  
  \frac{ 1}{ \Big(  \frac{ T}{t'} \, r'^2  \, + 
  T^2 \, log^2 \Big( \frac{t'}{ T'}\Big) \Big)^{(d-2)/2 } }  \ 
  & & \  \Big(  \frac{ T}{t'} \Big)^{\chi /2 } \ exp \Big( - \frac{ mr'^2}{4t'} \Big)
\EEA

 \section{ Energy-Momentum tensor, central charge }  
\par   It is a conjectural program to try to generalise to the above situation 
 the impressive achievements of conformal theories. Crucial concepts are 
  the stress tensor, which is not really a tensor since it does not transform 
  according to the homogeneous law of relativity but with an extra term proportional
   to the Schwarz derivative.  The proportionality coefficient, called up to a 
    rational number, central charge, measures the "number" of massless degrees of freedom 
    of the theory.  It is related to the celebrated trace anomaly and 
 can be calculated in two dimensions by various computations, such as 
   short distance expansions, partition functions evaluation 
 by various regularisations, 
    computation of finite size effect or spectrum of hamiltonians in 
 conformally flat geometries.
 An account of such computations is given in \cite{Capp89, Cost02}, as well as a  
 discussion of relationship between these concepts in $d>2 $. 
  Here the geometric point of view is important. For example the correct 
  bosonic massless Weyl invariant action in any dimension d should include 
  a term proportional to $R\  \phi^2  $ . We therefore give in appendix some 
  technicalities which will certainly reveal useful in this more  subttle 
   perspective.

%%%%%%%%%%%%%%%%%%%%%%%%%
\section{ Appendix 1 Riemann tensors}
Let us give some technicalities, recently considered  as fashionable  in 
membranes literature. 
If a non degenerate metric $G_{MN}= g_{\mu \nu} \bigoplus h_{mn} $ is block diagonal 
in term of coordinates $X^M =  (x^\mu , \, y^m )$, we have in Landau's conventions:  
\BEA  R_{\lambda \mu \sigma  \nu }  =  
  & & r_{\lambda \mu \sigma  \nu } (x, \, (y))\ +\, \frac{1}{4} h^{ab}\, ( 
  \partial_a  g_{\mu \sigma }\, \partial_b g_{\nu    \lambda }\, - 
  \partial_a  g_{\mu    \nu }\, \partial_b g_{\lambda \sigma }\,  ) \\   
  & &     \nonumber   
\EEA 
where $(y)$ means the coordinates $y^m $ are considered as parameters ie 
r is the Riemann tensor relative to 
 $g_{\mu , \nu }(x,y)$  but does not contain derivatives with respect to y. 
%%%%
We also need all mixed tensors such as: 
\BEA 
R_{\lambda \mu ,\, s  n } \ =& \frac{1}{4} &\  g^{\alpha \beta } \,  \Big(  
  \partial_s g_{\alpha  \mu  }\, \partial_n g_{\beta  \lambda }\,   -  
  \partial_n g_{\alpha  \mu  }\, \partial_s g_{\beta  \lambda }\, \Big)  \nonumber   \\  
 + & \frac{1}{4} &\  h^{ab}  \,               \Big(  
  \partial_{\mu }  h_{as} \partial_{\lambda } h_{bn} \, -  
  \partial_{\mu }  h_{an} \partial_{\lambda } h_{bs} \,   \Big) 
\EEA 
%%%%
\BEA 
 R_{ \mu  \nu } \   &=& r_{ \mu  \nu }(g(x,(y))) \nonumber \\  
 & &  -\frac{1}{2}  h^{ls} 
      \Big( \partial_l \partial_s  g_{\mu    \nu }\, -
 \Gamma_{ls}^a (h)  \, \partial_a  g_{\mu    \nu }\, \Big) \nonumber \\ 
 & &+ \frac{1}{2} g^{\alpha \beta } 
 (dg_{\mu \alpha }\cdot dg_{\nu  \beta } ) 
   -\frac{1}{4} \,  (dg_{\mu \nu } \cdot dlog(g) )\nonumber \\ 
  & & + \frac{1}{4} h^{ls}  h^{ab}\, 
   \partial_{\mu }  h_{as} \partial_{\nu }  h_{bl} \nonumber \\    
  & & - \frac{1}{2}  h^{ls}  \Big( 
  \partial_{\mu } \partial_{\nu }  h_{ls}  \,
 - \Gamma_{\mu \nu }^{\alpha }(g) \partial_{\alpha }  h_{ls} \Big) 
\EEA  
%%%%%%%%%
 Useful notations are: $ d_a log\  g \ =  g^{\mu \nu }\,  \partial_a g_{\mu \nu} $ and 
  $df\cdot dj \ = h^{ab}(x,y)\, \partial_a  f\, \partial_b j$ , 
 and  similarly 
 $\delta f\cdot \delta j =  g^{\mu \nu } \, \partial_{\mu }f \partial_{\nu }j $.
\BEA   R \ &=& r( g(x^{\mu }\, ,\,  (y)))\ + \ r(h((x),\, y^m )  \nonumber \\  
  & & -\frac{1}{4} (( dlog\  g \ \cdot  dlog\  g )+ 
         (\delta log\ h \cdot  \delta log\ h  ))         \nonumber \\  
  & & + \partial_a log\ g \, h^{mn} \, \Gamma_{mn}^a (h) \nonumber \\  
  & & + \partial_{\alpha } log \ h \ g^{\mu \nu }\, 
        \Gamma_{\mu \nu }^{\alpha } (g)  \nonumber \\  
  & & - g^{\mu \nu } h^{ab} ( \partial_a   \partial_b g_{\mu \nu }  
      + \partial_{\mu } \partial_{\nu } h_{ab} ) \nonumber \\  
  & & + \frac{3}{4} g^{\mu \nu } h^{ab} \, 
       \Big(  g^{\alpha \beta } \, \partial_{a} g_{\mu \alpha }  
             \partial_{b} g_{\nu \beta } \, 
      + h^{mn} \, \partial_{\alpha }  h_{am} \partial_{\beta } h_{bn} \Big)
\EEA

For a product of two manifolds, the scalar R is additive.  

 \section{ Appendix  2 Space and time in an accelerated laboratory}
 As a side application of mathematical methods used here we would like to 
  bring attention to the fact that such formulae 
  can be used for definition of  physical 
  space and time in an accelerated frame . 
  This could be useful for the study of expanding universe, black hole physics 
 (quantum melanodynamics), or of any accelerated matter system.  \\

 Physically, let's suppose we are in a rocket, or an errant planet,
 whose position of center of mass is $x= f(t) $ in a flat Minkovskian 
  space time endowed with coordinate system $(x,t) $. 
  
  For simplicity we'll write only one space coordinate; Coriolis forces 
  could be considered in a further step.   Classical arguments,
   given by Einstein, are that inside the moving object we dispose of a 
  {\bf physical } coordinate system $ x',\,  t' $  , and everything happens 
  similarly to what would happen in a locally inertial comoving frame.
  This means that we suppose the number $x' $ characterizes a material 
   point of our rocket, which is supposed (or kept ) fixed. 
   This is the case if $ {d^3 \, f \over d^3 \, t } =0 $ (we have in mind 
   the case of a rocket launched with constant acceleration, which should  
   be felt as equivalent to a gravitational field, and then 
following its trajectory at constant speed).  In formulae, we therefore 
have locally: 
\begin{eqnarray}
 dx' &=& {dx \, -\  v \ dt  \over \sqrt{ 1-\frac{v^2 }{c^2 }} } \\   
 dt' &=& {dt \, -\ \frac{v}{c^2 } \ dx \over \sqrt{1-\frac{v^2 }{c^2 }}}
\end{eqnarray} 
If speed $v$ is constant, this system leads to the celebrated Lorentz 
equations for $ x' $ and $t'$.  A point $x$ outside is seen from 
 the crew at  $x' = $ dilation factor $\times  (x-vt) $. The Minkovski 
 interval $ds^2 \, = (c^2 - v^2 ) \, dt^2 = c^2 d\tau^2 $ being conserved, 
 we have a proper time (at least at the center of mass $x'=0 $ ) 
   flowing slowlier according to: 
\begin{equation} 
  d\tau =dt' = \sqrt{ 1- { v^2 (t)  \over c^2 }} \ dt 
\end{equation}   

(This eq. follows from the above system if we have $dx' =0 $). \\ 
We would like to consider more carefully the situation where the speed
 is not a constant, and try to rigorously consider the above system as 
 made of p.d.e.'s for diffeomorphisms $ x'(t,x)  $ and $t'(t,x) $. \\ 
 We propose to consider as physical speed of a point labelled by $x'$ 
 the following quantity:\\  
 First, at any fixed $t$, invert $ x'=x'(t,x) $ into $ x= X(t, x') $ 
 (note this is different from inverting $(x,t) $ into $( x',t') $ ).
 Then define 
  \begin{equation} 
  v(t,x' ) := \frac{ \partial X}{ \partial t} (t,x') 
  \end{equation} 
  Therefore in the above p.d.e.'s system we have both time and space 
  dependent coefficients (that is necessary to avoid paradoxes) 
depending on: 
   \begin{equation} 
  v(t,x ) := \frac{ \partial X}{ \partial t} (t, x'(t,x)) 
  \end{equation} 
  
  Position of center of mass and proper time then appear 
  as boundary conditions:  
\begin{eqnarray}
 x'(t,f(t)) &=& 0   \\   
 t'(t,f(t)) &=& \tau (t)  
\end{eqnarray} 
  
%   \begin{eqnarray}
%  &=&    \nonumber \\   
%  &=&  
%   \end{eqnarray} 
   
%%%%%%%%%%%%%%%%%%%%%%%%%
\section{ Appendix 3 Geometric conformal theories}
 We would like to bring attention of the reader 
 to some geometric interpretation of so called conformal invariance 
 in Euclidean, critical, 2d statistical (field)  theories.
  "conformal transformations" are often considered as symmetries: 
  Any experienced mathematical physicists should take this with grains of salt,
  because a meromorphic function  $  Z= f(z) $ , e.g.   $= z^n $ is not 
   in general one to one and therefore is not an element of 
    the group of diffeomorphism of some projective  manifold. In 
    fact this is related to the rich theory of  ramified mappings, on which 
  physical mathematics  has also brought new enumerative results. Furthermore 
  geometric concepts are important here. Two  of these are  the concepts of 
  Riemann (also called normal or geodesic ) coordinates,  and  Weyl rescaling of 
   the Riemannian metric structure.  A way of understanding conformal invariance 
   is to consider the local diffeomorphism which expresses 
$  z' $, a  normal coordinate    after Weyl transformation  in terms 
 of $ z$ an old normal coordinate. This is explained and generalised to 
  higher dimension in  \cite{Capp89, Cost02, Eise40}

\noindent {\large\bf Acknowledgements}\\
  
 \noindent 
We thank  for discussions M. Dubois-Violette, 
C. Godr\`eche, H. Hilhorst, M. L\"uscher, Jean Bernard Zuber, 
P. Ruelle, as well as   Francesco Ravanini and Romano Prodi for kind hospitality
  and nice atmosphere   in Bologna. 

%%%%%%%%%%%%%%
 
   \end{document}